# A Proactive Design to Detect Denial of Service Attacks Using SNMP-MIB ICMP Variables

Yousef Khaled Shaheen
Department of Computer Science
Princess Sumaya University for Technology
Amman, Jordan
yousefpsut@icloud.com

Dr. Mohammad Al Kasassbeh
Department of Computer Science
Princess Sumaya University for Technology
Amman, Jordan
m.alkasassbeh@psut.edu.jo

*Abstract*— Denial of Service (DOS) attack is one of the most attack that attract the cyber criminals which aims to reduce the network performance from doing it's intended functions. Moreover, DOS Attacks can cause a huge damage on the data Confidentiality, Integrity and Availability. This paper introduced a system that detects the network traffic and varies the DOS attacks from normal traffic based on an adopted dataset. The results had shown that the adopted algorithms with the ICMP variables achieved a high accuracy percentage with approximately 99.6% in detecting ICMP Echo attack, HTTP Flood Attack, and Slowloris attack. Moreover, the designed model succeeded with a rate of 100% in varying normal traffic from various DOS attacks.

*Keywords—Cyber-attacks, availability, DOS attack, ICMP variables, ICMP Echo Attack, HTTP Flood Attack, Slowloris Attack*.

## I. INTRODUCTION

The wide use of the internet and the rapid increase of the communication and computer networks increase the cyber criminals' activities in attacking these networks and cause a catastrophic damage to them. Network security attacks are varied based on their effect on the network and the financial losses that they may cost the organization. DOS attack listed as one of the easiest attacks that can be launched with a huge impact on the network assets and cost the organizations a heavy losses. Many exhaustive researches have been done on the financial losses that DOS attack can cause. Ponemon Institute reported that, the average losses for 641 individuals are approximately equal to $1.5 million over the year of 2015 divided into five categories (Revenue Losses, Technical Support Costs, Operations Disruption, Lost User Productivity, and Damage to Information Technology Assets) [1]. Thus, many organizations aim to protect their networks from several attacks that can cost them heavy losses using different network security services. One of the common used security services is Intrusion Detection System (IDS), which is a security model that is designed to detect the abnormal and malicious traffic in real time or close to it. The idea behind DOS attack is to prevent a system from doing its intended functions and preventing the authorized users from accessing the system resources by injecting a flood of data to a specific target system. DOS attack can be categorized in two main techniques either by exploiting the vulnerabilities in the network servers, appliances and protocols, and exploiting a huge amount of spoofed source addresses. This paper introduced a new model to detect various DOS attacks by using a set of ICMP variables and an adopted dataset of these attacks. A set of algorithms such as, Meta, Lazy IBK, Bayes, RJ48 and Rule-Based were adopted to find which one of these algorithms is the most effective in detecting network anomalies.

This paper is organized as follows: Section 2 provides several related works in the area of using machine learning in detecting network anomalies, where the DOS attacks and SNMP-MIB dataset are illustrated in section 3. The proposed model that is used in this contribution is discussed in section 4. Section 5 discusses the experimental results of the adopted methodology. Finally, section 6 concludes the proposed model in this research paper.

## II. RELATED WORK

Most of the current researches have focused on detecting different network attacks using machine learning techniques. Many of these techniques has been introduced, tested, and evaluated. One of the most used techniques in detecting and analyzing network anomalies is SNMP-MIB data.

Al – Kasassbeh et al. [2] generated an effective dataset that solved the limitation resources in the previous datasets. The authors adopted a reliable SNMP-MIB dataset to investigate the SNMP for network attacks and anomalies detection. The authors collected SNMP-MIB data based on a set of Brute-Force attack and DOS attacks. The collected dataset is a reliable published dataset and it consists of 4998 records, where each record mapped to 34 MIB variables. The MIB groups are categorized as follows: TCP, UDP, IP, ICMP and Interface.

Al – Kasassbeh et al. [3] adopted a reliable method in detecting network anomalies based on SNMP-MIB dataset using machine learning techniques. They proved that the SNMP-MIB is an effective technique in detecting a large set of various DOS attacks using three algorithms categories; Random Forest, AdaboostMl, and MLP. The mentioned algorithms were applied on several MIB groups (TCP, UDP, IP, ICMP and Interface). The classified algorithms achieved a varied accuracy based on the group. The Random Forest algorithm achieved a high accuracy when it was applied on the IP group with a rate 100% and 99.93% when it was applied on the interface group.

Al – Kasassbeh [4] proposed a new hybrid approach to capture and detect the malicious traffic based on the collected dataset that is applied as an input to the Neural Network in order to predict the behaviour of input data. The proposed model achieved a high accuracy with a rate of 98.3% in capturing and detecting malicious traffic with minimal false negative rate.

Sharma et al. [5] proved that the volume-based analysis can't detect all types of network anomalies. The authors used some services such as, Simple Network Management Protocol (SNMP), Network Time Protocol (NTP), and Domain Name System (DNS) to analyse network anomalies. NfDump machine learning has been used to collect and capture the network packets.

Niyaz et al. [6] proposed a new scheme using a deep learning approach in order to improve the efficiency of network intrusion detection system. The authors evaluate the network anomalies detection based on NSL-KDD dataset, and achieved a high accuracy rate with minimal false alarms rates.

Suganya [7] adopted a new hybrid approach by combining two methods (Misuse based detection and network anomaly detection); to allow the built system to detect the malicious traffic without needing any previous knowledge of these traffic. The idea behind this approach is to characterize and differentiate the normal and malicious traffic, then the normal traffic will be applied under the process of anomaly detection. The authors achieved an efficient module in detecting DOS attack, as it found that the hybrid module is faster in detecting network attacks than the standalone methods (Misuse and Network anomaly detection). Moreover, the hybrid approach achieved a low false positive rate with high reliability.

Namvarasl and Ahmadzadeh [8] adopted a new intrusion detection approach based on two main approaches, the Simple Network Management Protocol and machine learning. The approached module is designed to detect DOS and DDOS attacks in real time and approximate to it. The authors designed their models based on three sub-modules, starting by collecting the MIB variables from a set of classifiers (C4.5, feature selection, and Ripper). The intrusion detection system has been set based on chosen variables to detect DoS attacks, where a dataset of 66 variables is mapped to 4 MIB groups (TCP, UDP, IP, and ICMP).

The previous modules are based on gathering their own dataset in order to come up with a scheme that detects network anomalies with the same concept of intrusion detection system. In this paper, 34 MIB variables were chosen and mapped to ICMP group in order to evaluate the accuracy of the proposed model.

III. DENIAL OF SERVICE ATTACKS AND SNMP-MIB DATASET

A. DOS Attacks

DOS attacks are one of the most attacks that attract intruders; since DOS is a form of attack on the service availability. NIST defines DOS attack as "A set of actions that comprise networks and its resources and preventing the authorized users from doing their intended functions".

DOS attacks compromise many network resources, these resources can be categorized as below:
- Network Bandwidth: network bandwidth relates to the channel capacity between the network appliances and the server, and the links capacity that connect a server to the global internet.
- System Resources: DOS attack can also target the system resources by overloading the server and crashing the handled services.
- Application Resources: in this category, DOS attack targets a specific application, such as a web server. This attack involves sending a flood of valid requests which consume the resources of a specific application.

This paper derived DOS attacks into 7 types based on the source that generated the flood traffic as follows:

SYN Spoofing attack is a basic flooding attack that targets the network server which is responsible of responding to TCP connection requests from network hosts. This attack aims to flood the server tables that manage and establish the connection between the server and any host in the network; which leads to denying any future request from legitimate users and prohibiting them from accessing the server.

SYN Spoofing attack depends mainly on the concept of three way handshaking, where the client starts establishing the connection by launching and sending TCP SYN packet to the network server that responds with a SYN ACK packet towards the client that replies with an ACK packet in return.

The idea behind SYN spoofing attack is to exploit the victim server behaviour by generating flood SYN packets with counterfeit source IP address. This attack represents in embedding a forged IP in SYN packet instead of the legitimate IP which leaves the server to respond with a SYN ACK packet to other clients in the internet cloud and reserving a space for the ACK packet in return, that leads the client to keep sending TCP SYN packet waiting for a SYN ACK and also leads the server to reply with SYN ACK packets to another client.

UDP Flooding attack represents in sending flood UDP packets to a specific port number of the victim server or system, which takes the specific port that handles a specific service down.

The fourth type of DOS attack represents in HTTP based attacks that are divided into two main categories, slowloris attack, and HTTP flood attack.

Slowloris attack engages by setting up several connections to a webserver, where in each established connection an incomplete request will be embedded in these connections and that doesn't include the terminating newline sequence, meanwhile the attacker keeps sending continuous header lines to keep the connection alive. After keeping the connection alive, the victim webserver keeps its connection open for any information that will complete the launched request, which eventually leaves the webserver with all its available channels consumed.

HTTP Flood attack revel the webserver with HTTP requests that established from several bots. This attack aims to consume the whole webserver resources and take it out of service. One of the common examples on this attack is exhausting the memory and consuming its capacity by overwhelming it with many tasks.

ICMP Echo attack depends mainly on the ping flood using echo request packets. The ease of using this attack and the reason behind considering it as a traditional attack is that the ICMP protocol is useful in network diagnostic, which leads most of network admins to control and restrict this protocol using different network security appliances such intrusion detection systems, intrusion prevention systems or firewalls. However, this protocol is also critical to some networks such TCP/IP networks. Intruders in this attack generate a huge volume of ICMP packets toward the victim server which utilize the link bandwidth that leads other users to face difficulties while reaching to the victim server.

Table 1 classifies the dataset records according to the related attacks.

TABLE I. DATASET RECORDS ACCORDING TO RELATED ATTACKS

| No. | Traffic Label | Traffic Count |
|---|---|---|
| 1 | Normal | 600 |
| 2 | ICMP-Echo Attack | 632 |
| 3 | TCP-SYN Attack | 960 |
| 4 | UDP Flood Attack | 773 |
| 5 | HTTP Flood Attack | 573 |
| 6 | Slowloris Attack | 780 |
| 7 | Slowpost Attack | 480 |
| 8 | Brute Force Attack | 200 |

### B. Simple Network Management Protocol (SNMP)

SNMP is an application layer protocol that allows the user to monitor, analyse and manage network traffic. SNMP protocol divides into three versions that vary in features; SNMPv1 and SNMPv2 are known as SNMP community, where SNMPv3 known as SNMP security, the only difference between these three versions is that SNMPv3 designed with advanced security features. Fig.1 illustrates the network management architecture.

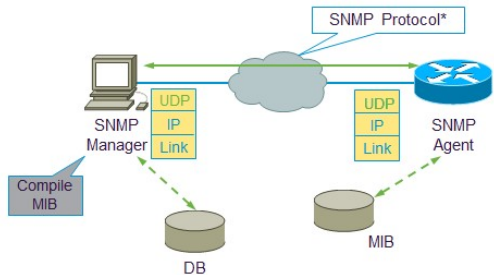

Fig.1. Network Management Architecture [9]

Fig.1. shows that SNMP network model is divided into two main subsystems; the SNMP Manager and the SNMP Agent. The SNMP Manager is a personal computer that is designed and configured to pull the data from SNMP Agent. SNMP Manager is designed to provide solution for a set of faults and categories such as, fault monitoring, performance monitoring, configuration control and the security control.

SNMP Agent plays the main role in network management model, by collecting the required data from the network and stores them in a database called the Management Information Base (MIB). SNMP Agent is embedded on the required device, where it responds and exchanges the requests and the actions from the SNMP Manager using SNMP protocol.

## IV. PROPOSED MODEL

This part is divided into three sections, starting with a brief description of the used dataset. The second section provides a fully explanation of the machine learning classifiers that is used to classify the dataset and create a decision if either a normal traffic or a malicious traffic. The last part provides a summary of feature selection techniques that are used in the module to evaluate the efficiency of applying these features on the ICMP variables.

### A. SNMP-MIB Data

In this research paper, (Al – Kasassbeh et al. 2016) SNMP-MIB dataset was used for testing and implementing this paper approach. The dataset was built from almost 5000 records that are related to seven main types of attacks (ICMP Echo, TCP-SYN, UDP flood, HTTP flood, Slowpost, Slowloris and Brute Force attacks). The set of attacks were detected using a set of variables that are included in the dataset. The traffic prediction is based on ICMP group. Most of network traffic deals with ICMP protocol to ensure the best packet delivery by comparing the number of sent and received packets. Six MIB variables were selected for this group as follows.

- The icmpOutMsgs (iOM) variable indicates the total count of attempt ICMP sending messages.
- The icmpInMsgs (iIM) variable is an indicator of the total number of ICMP received messages.
- The icmpOutDestUnreachs (iOU) variable indicates the total amount of unreachable ICMP messages sent at the destination.
- The icmpInDestUnreachs (iIU) variable is an indicator of the total count of unreachable ICMP messages at the destination.
- The icmpInEchos (iIE) variable indicates the total ICMP number request packets received.
- The icmpOutEchos (iOE) variable indicates the total ICMP number reply packets received.

### B. Machine Learning Classifiers

The idea behind using classifiers in network anomaly detection system is to analyze and classify the corresponding traffic. In this paper, five classifiers were applied on the adopted dataset as follows.

- Meta Bagging classifier was presented by Efron Tibshirani. Bagging is a Meta bootstrap algorithm that trains every single classifier randomly of the original dataset to generate and form a final prediction. The bagging classifier is divided into two categories based on the dataset subset; if the dataset subsets are drawn randomly, then it called pasting. While if the dataset subsets are drawn with replacement, then it called bagging.

- Lazy classifier is known as an algorithm or a system that trains and generalizes the records in the dataset after the system receives queries. Lazy IBK classifier is applied on the adopted dataset, since it proved its efficiency when applying it on large datasets with various attributes.

- J48 classifier is an implementation branch of tree classifier family that is also called C4.5. J48 algorithm was introduced and developed by Ross Quinlan. The process of attribute selection is done over top down induction of decision trees and then uses information theory key concepts in order to select the best attribute.
- Rule-based classifier is one of the most commonly used algorithms in artificial intelligence science, due to the high accuracy provided results. The role of this classifier is using a set of rules in order to generate several choices. Rule-based classifier falls into two characteristics, the mutually exclusive rule, where each record in the dataset is covered mostly by one rule. In the addition, exhaustive rule occurs when each record in the dataset is covered at least by one rule.
- Bayes classifier is also known as Naïve Bayes, this classifier was developed by Thomas Bayes. The role of this classifier is conditional probability; which is the probability of something to happen based on something else has already occurred. Bayes classifier executes the probabilities for each class of the dataset, where the highest probability rate is the most occurring class.

*C. Attribute Selection*

Attribute selection was adopted to improve the proposed model by reducing the factors count in the dataset which the designed model needed it for testing and learning stages. Thus, attribute selection neglected the irrelevant fields while providing accurate results at the same time.

Attribute selection techniques fall in three main categories as follows.

- The filter technique operates by selecting features based on the features earlier scores in various statistical tests corresponding to the outcome variables. The strength of this technique is obvious when applying it on large datasets.
- The wrapper technique looks through the feature space and uses the algorithm to find the best attribute set. The searching method of wrapper technique can be in several directions (forward, backwards, or bidirectional). The strength in wrapper technique relates to the efficient results because of the complexity of this method, as it participates in the selection process.
- The hybrid approach combines both the filter and the wrapper technique which results into a complex feature selection technique.

The Filter and the wrapper techniques were used in order to compare the accuracy of the generated results, where for filter technique two methods were selected the infoGain and the ReleifF, and for the wrapper technique correlation-based method was selected.

InfoGain, ReleifF, and Correlation-based are attribute evaluators that are used in WEKA machine learning tool. InfoGain finds out the most useful attribute for prejudiced between the various classes to be used. Moreover, InfoGain determines the best split to be chosen; the more accurate split is the one that has the high value.

ReleifF attribute evaluator is an effective method of attribute ranking. The role behind selecting is the more important attribute is based on the algorithm output; the more positive number means the more important attribute, where the output is a number that varies between -1 and 1. The attribute weight is continuously updated through the process. Three samples are selected and recognized respectively, a selected sample from the dataset, the closest neighbouring sample that belongs to the same class in the dataset, and the closest neighbouring sample in different class in the dataset. The attribute weight is affected by any change that can be done on any attribute value which also could be responsible for the class change.

Correlation-based evaluator is based on finding the correlation between two related features by evaluating the correlation coefficient. The attribute can be redundant by either deriving it from another set of attributes or if it's related to some other attributes. So that, to consider an attribute as a good attribute it should be highly correlated to the attributes class and not highly correlated to any other attributes. Table 2 shows the ICMP variable ranks when they were applied under the attributes selection factors.

TABLE II. DATASET RECORDS ACCORDING TO RELATED ATTACKS

| Attribute Selection Factor | Top 4 ICMP Variables Ranking | Top 3 ICMP Variables Ranking |
|---|---|---|
| ReliefF | iIU | iIU |
| | iOM | iOM |
| | iIE | |
| | iOE | iIE |
| InforGain and Correlation-Based | iOU | iOU |
| | iIE | |
| | iOE | iIE |
| | iOM | iOE |

The search method in this paper was done by using the ranker searching method, which ranked the attributes based on their evaluations from the highest importance to the lowest one.

*D. Evaluation Metrics*

The performance of the proposed model was measured using a set of well-known parameters such as accuracy, precision, and recall. The classifiers performance was measured based on the confusion matrix as follows.

| | Predicted Class | |
|---|---|---|
| Actual Class | Positive | Negative |
| Positive | TP | FP |
| Negative | FN | TN |

TABLE III. CONFSUSION MATRIX

The true positive (TP) rate indicates the rate of the correct predictions of the positive traffic proverbs. False positive (FP) rate indicates to the proportion of negative packets that are positive packets. The true negative (TN) rate is an indication of the total number of negative traffic that classified correctly as negative, where the false negative (FN) rate shows the total number of positive traffic that classified incorrectly as negative traffic.

The precision rate represents the ratio of the total correct predictions of the positive traffic proverbs to the total count of irrelevant and relevant traffics. The recall accuracy rate represents the attribution of the correct prediction rate of the positive traffic instances to the total count of relevant traffic instances.

Finally the accuracy rate takes all confusion matrix parameters into its calculation to measure the correctly classified traffic instances. Precision, recall and accuracy formulas are showing below respectively.

$$\text{Precision} = \frac{TP}{TP+FP} \quad (1)$$

$$\text{Recall} = \frac{TP}{TP+FN} \quad (2)$$

$$\text{Accuracy} = \frac{TP+TN}{TP+FP+TN+FN} \quad (3)$$

The F-Measure metric was measured in this research paper, where its calculation depends on the following formula.

$$\text{F-Measure} = 2 \, x \, \frac{Precision \, x \, Recall}{Precision+Recall} \quad (4)$$

## V. EXPERIMENTAL RESULTS

The results and the performance were evaluated over Weka machine learning, version 3.8 that is running over Inter® core[TM] i5, 64-bits system with 4 GB RAM running on windows 10.

The results of the proposed model depended on the MIB dataset that mentioned earlier in section 4. The classification techniques were then applied with each group separately. At the end, the attribute selection methods were used to evaluate the accuracy of the proposed model by reducing the factor number in the dataset. Table 4 shows the weighted accuracy rate for all classes based on the ICMP-Echo group and the selected classifiers.

TABLE IV. CLASSIFIERS ACCURACY FACTORS AVERAGE WEIGHT

| Classifiers | Accuracy Factors Average Weight | | | | |
|---|---|---|---|---|---|
| | TP Rate | FP Rate | Precision | Recall | F-Measure |
| Bayes | 0.864 | 0.014 | 0.935 | 0.864 | 0.879 |
| Lazy-IBK | 0.867 | 0.026 | 0.895 | 0.867 | 0.872 |
| Met -Bagging | 0.871 | 0.029 | 0.906 | 0.871 | 0.874 |
| Rules-Based | 0.867 | 0.026 | 0.895 | 0.867 | 0.872 |
| RJ.48 | 0.868 | 0.026 | 0.896 | 0.868 | 0.872 |

Fig.2. illustrates the performance of the classifiers that were used in the proposed model in terms of F-Measure rates and based on the ICMP variables.

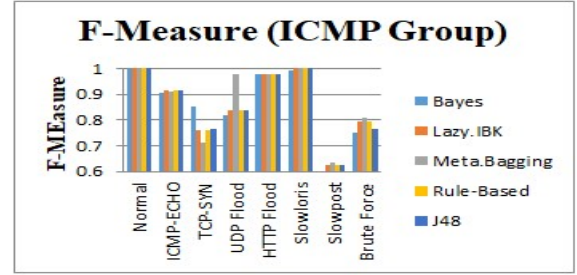

Fig.2. F-Measure Results of All ICMP Group

From Fig.2 it was found that the F-Measure values of all classifiers are efficient for normal traffic, HTTP flood attack, and slowloris attack. Moreover, Meta Bagging classifier achieved a high performance in identifying UDP flood attack.

Figs.3, 4 illustrate the performance of the classifiers in terms of F-Measure based on top 4 and top 3 ICMP variables that were selected using attribute evaluators.

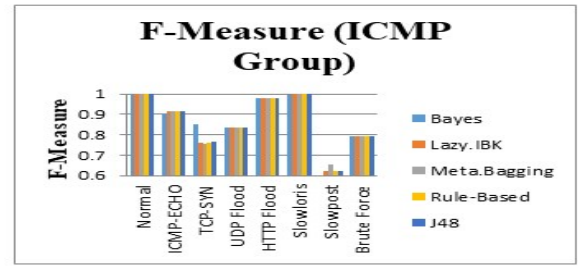

Fig.3. F-Measure Results with Top 4 ICMP Variables – ReliefF Evaluator

Fig.3 had shown that the F-measure results of all classifiers are effcient for normal traffic, ICMP Echo attack, HTTP flood attack, and Slowloris attack. On the other hand, the F-Measures of the other classifiers were fragile in detecting the rest types of DOS attack.

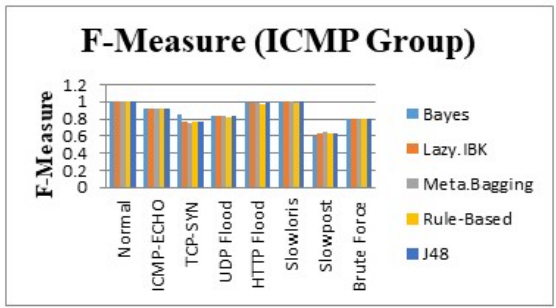

Fig.4. F-Measure Results with Top 3 ICMP Variables – ReliefF Evaluator

From fig.4 it was found that all classifiers achieved a high F-Measure rates for normal traffic, HTTP flood attack, ICMP Echo attack, and slowloris attack. However, all classifiers werent efficient in detecting the rest types of attacks.

InfoGain and correlation evaluator selectors achieved equalized F-Measure values when applied them on top 3 and top 4 ICMP variables as shown in fig.5, 6.

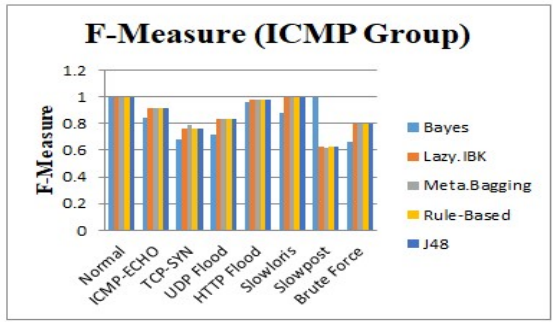

Fig.5. F-Measure results with Top 4 ICMP Variable – InfoGain and Correlation Evaluators

Fig.5 shows that all classifiers achieved high F-Measure rates for normal traffic, HTTP flood attack, ICMP Echo attack, and Slowloris attack. Moreover, the Bayes classifier in the Slowpost attack achieved a high performance in identifying this type of attack only. However, all classifiers are not efficient in detecting the remaining types of DOS attack.

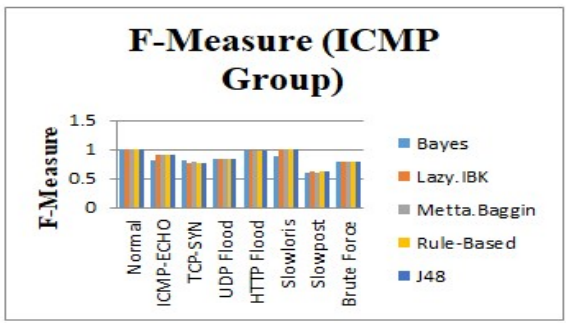

Fig.6. F-Measure results with Top 4 ICMP Variable – InfoGain and Correlation Evaluators

Fig.6 shows that all classifiers achieved high F-measure rates for normal traffic, HTTP flood attack, ICMP Echo attack, and Slowloris attack. However, Bayes classifier failed in detecting ICMP Echo attack.

## VI. CONCLUSION

Data filtering becomes essential to protect the local and remote networks from different types of attack that harm the sensitive data and cost the organizations heavy losses. So that, many methods were introduced to detect network anomalies in order to keep network structure running normally without any disturbance and data disruption. In this paper, it was found that the ICMP group with the adopted classifiers wasn't efficient in detecting all DOS attacks. Moreover, reducing the count of ICMP variables varied in their performance when detecting these attacks. However, the designed model achieved an efficient performance in detecting some attacks such as, ICMP Echo attack, HTTP flood attack, and slowloris attack.

## VII. REFERENCES

[1] P. Institute, "2015 Cost of Data Breach Study: Global Analysis," Ponemon Institute, 2015.

[2] M. Alkasassbeh, G. Al-Naymat and E. Hawari, "Towards Generating Realistic SNMP-MIB Dataset for Network Anomaly Detection," *International Journal of Computer Science and Information Security (IJCSIS),* vol. 14, no. 9, pp. 1161-1185, 2016.

[3] M. Alkasassbeh, G. Al-Naymat and E. Hawari, "Using machine learning methods for detecting network anomalies within SNMP-MIB dataset," *International Journal of Wireless and Mobile Computing,* vol. 15, no. 1, pp. 67-76, 2018.

[4] M. Alkasassbeh, "A Novel Hybrid Method for Network Anomaly Detection Based on Traffic Prediction and Change Point Detection".

[5] R. Sharma, A. Guleria and R. K. Singla, "Characterizing Network Flows for Detecting DNS, NTP, and SNMP Anomalies," in *Intelligent Computing and Information and Communication*, 2018.

[6] Q. Niyaz, W. Sun, A. Y. Javaid and M. Alam, "A Deep Learning Approach for Network Intrusion Detection System".

[7] R. Suganya, "Denial-of-Service Attack Detection Using Anomaly with Misuse Based Method," *IJCSNS International Journal of Computer Science and Network Security,* vol. 16, no. 4, pp. 124-128, 2016.

[8] S. NAMVARASL and M. AHMADZADEH, "A Dynamic Flooding Attack Detection System Based on Different Classification Techniques and Using SNMP MIB Data," *International Journal of Computer Networks and Communications Security,* vol. 2, no. 9, p. 279–284, 2014.

[9] "Cisco," Cisco, [Online]. Available: http://www.cisco.com. [Accessed 3 April 2019].